\pdfoutput=1
\documentclass[sigconf]{acmart}
\makeatletter
\gdef\@copyrightpermission{
 \begin{minipage}{0.3\columnwidth}
  \href{https://creativecommons.org/licenses/by/4.0/}{\includegraphics[width=0.90\textwidth]{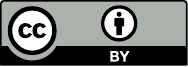}}
 \end{minipage}\hfill
 \begin{minipage}{0.7\columnwidth}
  \href{https://creativecommons.org/licenses/by/4.0/}{This work is licensed under a Creative Commons Attribution International 4.0 License.}
 \end{minipage}
 \vspace{5pt}
}
\def\@ACM@checkaffil{
    \if@ACM@instpresent\else
    \ClassWarningNoLine{\@classname}{No institution present for an affiliation}%
    \fi
    \if@ACM@citypresent\else
    \ClassWarningNoLine{\@classname}{No city present for an affiliation}%
    \fi
    \if@ACM@countrypresent\else
        \ClassWarningNoLine{\@classname}{No country present for an affiliation}%
    \fi
}

\usepackage{enumitem}
\usepackage{multirow}
\usepackage{bm}
\usepackage{subfigure}
\usepackage{textcomp}%

\AtBeginDocument{%
  }

\copyrightyear{2024}
\acmYear{2024}
\setcopyright{rightsretained}
\acmConference[RecSys '24]{18th ACM Conference on Recommender Systems}{October 14--18, 2024}{Bari, Italy}
\acmBooktitle{18th ACM Conference on Recommender Systems (RecSys '24), October 14--18, 2024, Bari, Italy}\acmDOI{10.1145/3640457.3688120}
\acmISBN{979-8-4007-0505-2/24/10}

\begin{document}

\title{Improving the Shortest Plank: Vulnerability-Aware Adversarial Training for Robust Recommender System}

\author{Kaike Zhang}
\affiliation{%
  \institution{CAS Key Laboratory of AI Safety, Institute of Computing Technology, Chinese Academy of Sciences}
  \country{ }
}
\affiliation{%
  \institution{University of Chinese Academy}
  \country{of Sciences, Beijing, China}
}
\email{zhangkaike21s@ict.ac.cn}

\author{Qi Cao}
\affiliation{%
  \institution{CAS Key Laboratory of AI Safety, Institute of Computing Technology, Chinese Academy of Sciences,}
  \country{Beijing, China}
}
\email{caoqi@ict.ac.cn}
\authornote{Corresponding author}

\author{Yunfan Wu}
\affiliation{%
  \institution{CAS Key Laboratory of AI Safety, Institute of Computing Technology, Chinese Academy of Sciences}
  \country{ }
}
\affiliation{%
  \institution{University of Chinese Academy}
  \country{of Sciences, Beijing, China}
}
\email{wuyunfan19b@ict.ac.cn}

\author{Fei Sun}
\affiliation{%
  \institution{CAS Key Laboratory of AI Safety, Institute of Computing Technology, Chinese Academy of Sciences,}
  \country{Beijing, China}
}
\email{sunfei@ict.ac.cn}

\author{Huawei Shen}
\affiliation{%
  \institution{CAS Key Laboratory of AI Safety, Institute of Computing Technology, Chinese Academy of Sciences,}
  \country{Beijing, China}
}
\email{shenhuawei@ict.ac.cn}

\author{Xueqi Cheng}
\affiliation{%
  \institution{CAS Key Laboratory of AI Safety, Institute of Computing Technology, Chinese Academy of Sciences,}
  \country{Beijing, China}
}
\email{cxq@ict.ac.cn}

\renewcommand{\shortauthors}{Kaike Zhang et al.}

\begin{abstract}
Recommender systems play a pivotal role in mitigating information overload in various fields. Nonetheless, the inherent openness of these systems introduces vulnerabilities, allowing attackers to insert fake users into the system's training data to skew the exposure of certain items, known as poisoning attacks. Adversarial training has emerged as a notable defense mechanism against such poisoning attacks within recommender systems. Existing adversarial training methods apply perturbations of the same magnitude across all users to enhance system robustness against attacks. Yet, in reality, we find that attacks often affect only a subset of users who are vulnerable. These perturbations of indiscriminate magnitude make it difficult to balance effective protection for vulnerable users without degrading recommendation quality for those who are not affected. To address this issue, our research delves into understanding user vulnerability. Considering that poisoning attacks pollute the training data, we note that the higher degree to which a recommender system fits users' training data correlates with an increased likelihood of users incorporating attack information, indicating their vulnerability. Leveraging these insights, we introduce the \textbf{V}ulnerability-aware \textbf{A}dversarial \textbf{T}raining (\textbf{VAT}), designed to defend against poisoning attacks in recommender systems. VAT employs a novel vulnerability-aware function to estimate users' vulnerability based on the degree to which the system fits them. Guided by this estimation, VAT applies perturbations of adaptive magnitude to each user, not only reducing the success ratio of attacks but also preserving, and potentially enhancing, the quality of recommendations. Comprehensive experiments confirm VAT's superior defensive capabilities across different recommendation models and against various types of attacks.

\end{abstract}

\ccsdesc[500]{Information systems~Recommender systems}
\ccsdesc[500]{Security and privacy~Social network security and privacy}

\keywords{Robust Recommender System, Adversarial Training, Poisoning Attack}


\maketitle

\section{INTRODUCTION}
\begin{figure}
    \centering
    \subfigure[Number of unaffected versus affected users across different attacks.]{
    \includegraphics[width=0.48\textwidth]{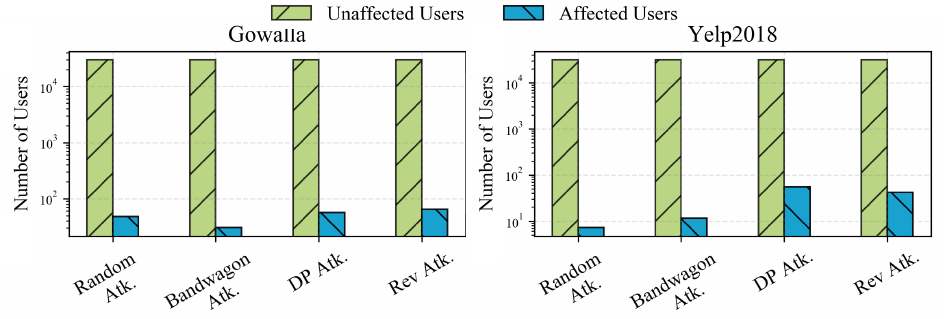}
    \label{fig:intro_attack}
    }
    \subfigure[Adversarial training as a defense mechanism against attacks.]{
    \includegraphics[width=0.48\textwidth]{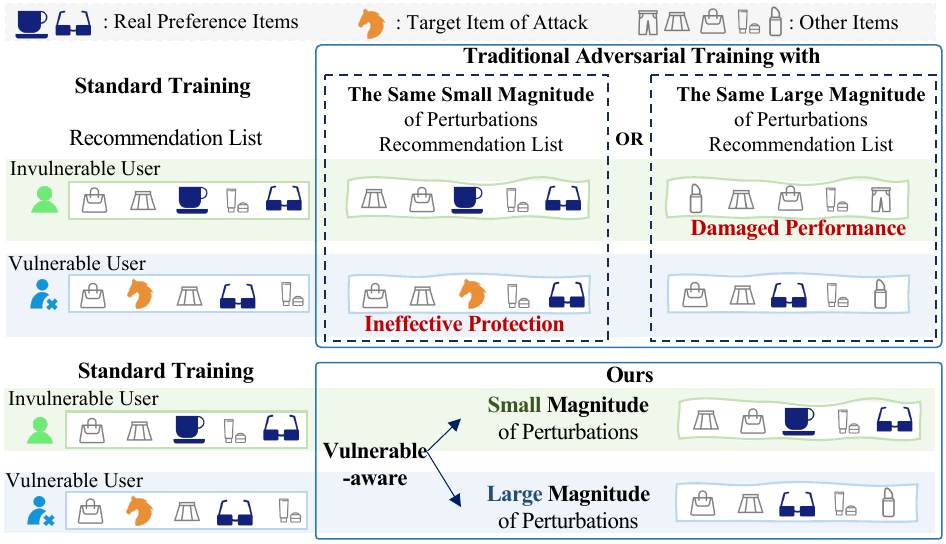}
    \label{fig:motivation}
    }
    \caption{(a) Illustrates that only a minority of users are affected by a given attack. (b) Demonstrates that applying the same magnitude of perturbations can lead to damaged performance for users not vulnerable to attacks or fail to effectively protect those who are vulnerable.}
\end{figure}

Recommender systems have become essential tools for managing the exponential growth of information available online. Collaborative Filtering (CF) is particularly notable among the various techniques employed in these systems~\cite{koren2009matrix, he2020lightgcn}. CF is deployed across diverse domains, from e-commerce platforms~\cite{smith2017two} to content streaming services~\cite{gomez2015netflix}, significantly enhancing user experience by offering personalized suggestions. However, the openness of recommender systems also makes them vulnerable to attacks where attackers inject fake users into the system's training data, known as poisoning attacks. These attacks aim to manipulate the exposure of targeted items~\cite{huang2021data, tang2020revisiting}, which may not only degrade the user experience but also pose a threat to the long-term sustainability of recommender systems~\cite{zhang2023robust, zhang2024lorec}.

Existing methods for combating poisoning attacks within recommender systems can generally be divided into two main approaches~\cite{zhang2023robust}: (1) detecting and removing malicious users from the dataset~\cite{chung2013beta, yang2016rescale, zhang2020gcnbased, zhang2024lorec}, and (2) developing robust models through adversarial training~\cite{he2018adversarial, li2020adversarial, wu2021fight, ye2023towards}. Detection-based approaches inspect entire datasets to eliminate malicious users, often relying on supervised data~\cite{chung2013beta, yang2016rescale} or specific assumptions regarding the characteristics of malicious users~\cite{zhang2020gcnbased}. These approaches may fail when characteristics of practical attacks deviate from predefined criteria. On the other hand, adversarial training improves model robustness by introducing perturbations into the embeddings of users and items during the training phase, utilizing a ``min-max'' strategy to minimize risks under the worst-case attacks~\cite{he2018adversarial, li2020adversarial, chen2023adversarial}, providing a more general and effective defense without the need for prior knowledge.

Considering the above advantages, this paper focuses the paradigm of adversarial training. Existing adversarial training methods typically apply the same magnitude of perturbations to all users. However, in practical scenarios, only a minority of users might be affected by attacks, as illustrated in Figure~\ref{fig:intro_attack}~\cite{huang2021data, tang2020revisiting, li2022revisiting}. To protect users against attacks, introducing the same large-magnitude perturbations across all users may inevitably impair the experience for those not vulnerable to attacks. Conversely, applying the same small-magnitude perturbations results in inefficient protection, as shown in the top part of Figure~\ref{fig:motivation}. Thus, applying the same magnitudes of perturbations across all users may not be the most effective way, resulting in a trade-off between recommendation performance and effective protection against poisoning attacks.

To address this issue, we propose user-adaptive-magnitude perturbations in adversarial training. On the one hand, we prioritize identifying users vulnerable to attacks to ensure their performance remains robust through sufficiently large perturbations. On the other hand, for users deemed less vulnerable, we propose reducing the magnitude of adversarial perturbations, as depicted in the bottom part of Figure~\ref{fig:motivation}. Unfortunately, from a defensive standpoint, without details about the attackers' targets, it is challenging to assess which users are affected by attacks, making precise identification of user vulnerability difficult.

Given these challenges, we explore alternative indicators to estimate user vulnerability. Through extensive experiments, we find that a user's vulnerability to attacks changes as the recommendation system undergoes training, indicating that this vulnerability is fluctuant. This insight leads us to further examine the link between a user's vulnerability and the model's training process. Considering the nature of poisoning attacks---where training data is polluted to mislead the recommender system from fitting the user's real preferences---we pose the question: Is there a correlation between a user's vulnerability and their degree of fit within the recommender system? Our empirical analysis yields a noteworthy discovery: \textit{users with a higher degree of fit within the recommender system face a higher risk of being affected by attacks}. This finding is intuitive in the context of poisoning attacks, as users with a higher degree of fit are also more likely to capture the malicious patterns in poisoned data, placing them at a greater risk.

Based on this observation, we propose a \textbf{V}ulnerability-Aware \textbf{A}dversarial \textbf{T}raining (\textbf{VAT}) method to enhance the robustness of recommender systems. VAT follows the established adversarial training paradigm in recommender systems~\cite{he2018adversarial}, which introduces adversarial perturbations to user and item embeddings during the training phase. To protect users who are vulnerable to attacks while preserving the performance of those who are not, we implement a vulnerability-aware function. This function estimates users' vulnerabilities based on the degree to which the recommender system fits them. Following this assessment, VAT applies user-adaptive magnitudes of perturbations to the embeddings. In this way, VAT can both diminish the success ratio of attacks and maintain recommendation quality, thereby avoiding the trade-off suffered by traditional adversarial training methods. 

Our extensive experiments across multiple recommendation models and various attacks consistently show that VAT significantly enhances the robustness of recommender systems (reducing the average success ratio of attacks by 21.53\%) while avoiding a decline in recommendation performance (even improving the average recommendation performance of the backbone model by 12.36\%). The pivotal contributions of our work are as follows:
\begin{itemize}[leftmargin=*]
    \item Through extensive empirical analysis, we interestingly find that ``users with a higher degree of fit within the recommender system are at a higher risk of being affected by attacks''.
    \item Building on these insights, we introduce a novel vulnerability-aware adversarial training method, i.e., VAT, applying user-adaptive magnitudes of perturbations based on users' vulnerabilities.
    \item Our comprehensive experiments confirm the effectiveness of VAT in resisting various attacks while maintaining recommendation quality, as well as demonstrating its adaptability across various recommendation models.
\end{itemize}

\section{RELATED WORK}
This section briefly reviews the research on collaborative filtering, poisoning attacks, and robust recommender systems.

\subsection{Collaborative Filtering}
Collaborative Filtering (CF) is a foundational technique in modern recommender systems, widely recognized and applied across the field~\cite{covington2016deep, ying2018graph, he2020lightgcn}. Its core assumption is that users with similar preferences tend to share comparable opinions and behaviors~\cite{koren2021advances}, which can be leveraged to predict future recommendations. Among CF methods, Matrix Factorization (MF) is particularly prominent, as it models latent user and item embeddings by decomposing the observed interaction matrix~\cite{koren2009matrix}. The integration of deep learning technologies has led to the emergence of neural CF models that aim to uncover more complex patterns in user preferences~\cite{wang2015collaborative, he2017neural, liang2018variational, he2018outer}. More recently, the advent of Graph Neural Networks~\cite{wu2020comprehensive} has facilitated the development of graph-based CF models such as NGCF~\cite{wang2019neural} and LightGCN~\cite{he2020lightgcn}, achieving notable success in enhancing recommendation tasks. Despite these advancements, these systems remain susceptible to poisoning attacks, posing significant challenges to their robustness~\cite{zhang2023robust}.

\subsection{Poisoning Attacks in Recommender System}
Poisoning attacks in recommender systems involve injecting fake users into the training data to manipulate the exposure of certain items. Early works focus on rule-based heuristic attacks. These methods typically construct profiles for fake users through heuristic rules~\cite{lam2004shilling, burke2005limited, mobasher2007toward, seminario2014attacking}. The Random Attack~\cite{lam2004shilling} involves fake users engaging with targeted items along with a random selection of other items. Conversely, the Bandwagon Attack~\cite{burke2005limited} crafts fake users' interactions to include both targeted items and those chosen for their popularity. As attacks have advanced, more recent contributions have adopted optimization-based approaches to generate fake user profiles~\cite{li2016data, huang2021data, wu2021triple, tang2020revisiting, li2022revisiting, chen2022knowledge, qian2023enhancing, wang2023poisoning, chen2023dark, huang2023single}. For instance, the Rev Attack~\cite{tang2020revisiting} frames the attack as a bi-level optimization challenge, tackled using gradient-based techniques. The DP Attack~\cite{huang2021data} targets the attack of deep learning recommender systems.

\subsection{Robust Recommender System}
Mainstream strategies for enhancing the robustness of recommender systems against poisoning attacks typically fall into two groups~\cite{zhang2023robust}: (1) detecting and excluding malicious users~\cite{chirita2005preventing, chung2013beta, yang2016rescale, zhang2020gcnbased, zhang2024lorec, liu2020recommending, zhang2014hhtsvm}; (2) developing robust models through adversarial training~\cite{he2018adversarial, chen2019adversarial, li2020adversarial, wu2021fight, ye2023towards, chen2023adversarial}.

Detection-based methods aim to either identify and remove potential fake users from the dataset~\cite{chung2013beta, zhang2014hhtsvm, yang2016rescale, liu2020recommending} or mitigate the impact of malicious activities during training~\cite{zhang2020gcnbased, zhang2024lorec}. These approaches often depend on specific assumptions about attack patterns~\cite{chung2013beta, zhang2020gcnbased} or supervised data regarding attacks~\cite{zhang2014hhtsvm, yang2016rescale, zhang2020gcnbased, zhang2024lorec}. Among these, LoRec~\cite{zhang2024lorec2} leverages the expansive knowledge of large language models to enhance sequential recommendations, overcoming the limitations associated with specific knowledge in detection-based strategies. However, its applicability is limited to sequential recommender systems and is hard to extend to CF.

In contrast, mainstream adversarial training methods for recommender systems, such as Adversarial Personalized Ranking (APR)~\cite{he2018adversarial}, introduce adversarial perturbations at the parameter level during training~\cite{he2018adversarial, li2020adversarial, ye2023towards, chen2023adversarial}. This methodology adopts a ``min-max'' optimization strategy, aiming to minimize recommendation errors while maximizing the impact of adversarial perturbations~\cite{zhang2024lorec}. This approach requires the model to maintain recommendation accuracy under the worst attacks, within a predefined perturbation magnitude. In practice, however, only a minority of users may be affected by attacks~\cite{huang2021data, tang2020revisiting}. Adversarial training that imposes the same large-magnitude perturbations prepares every user for the worst attacks, potentially degrading the experience for users unaffected by attacks. Conversely, the same small-magnitude perturbations offer insufficient protection for vulnerable users. Thus, there is a critical need for a technique that offers targeted protection to vulnerable users while preserving the recommendation performance of those who are not vulnerable.

\section{PRELIMINARY}
This section mathematically formulates the task of collaborative filtering and adversarial training for recommender systems.

\textbf{Collaborative Filtering}. Collaborative filtering (CF) methods are extensively used in recommender systems. Following~\cite{su2009survey, he2020lightgcn}, we define a set of users $\mathcal{U} = \{u\}$ and a set of items $\mathcal{I} = \{i\}$. We aim to learn latent embeddings $\bm{P} = [\bm{p}_u \in \mathbb{R}^d]_{u \in \mathcal{U}}$ for users and $\bm{Q} = [\bm{q}_i \in \mathbb{R}^d]_{i \in \mathcal{I}}$ for items. Subsequently, we employ a preference function $f: \mathbb{R}^d \times \mathbb{R}^d \rightarrow \mathbb{R}$, which predicts user-item preference scores, denoted as $\hat{r}_{u,i} = f(\bm{p}_u ,\bm{q}_i)$.

\textbf{Adversarial Training for Recommender Systems}.  
Adversarial training approaches for recommender systems, particularly within the Adversarial Personalized Ranking (APR) framework~\cite{he2018adversarial}, incorporate adversarial perturbations at the parameter level throughout the training process. The original loss function of the recommender system is represented as $\mathcal{L}(\Theta)$, with $\Theta = (\bm{P}, \bm{Q})$ indicating the system's parameters. These adversarial training methods introduce perturbations $\Delta$ directly to the parameters as follows:
\begin{equation}
\label{eq:at}
    \begin{aligned}
        \mathcal{L}_{\mathrm{AT}}(\Theta) =& \mathcal{L}(\Theta) + \lambda \mathcal{L}(\Theta + \Delta^{\mathrm{AT}}), \\
        \mathrm{where} \quad \Delta^{\mathrm{AT}} =& \arg \max_{\Delta,\, \Vert \Delta \Vert \leq \epsilon} \mathcal{L}(\Theta+\Delta),
    \end{aligned}  
\end{equation}
where $\epsilon > 0$ controls the magnitude of perturbations, and $\lambda$ denotes the adversarial training weight. 
In practice~\cite{he2018adversarial}, for an interaction $(u, i)$, the specific adversarial perturbation is given by
\begin{equation}
    \begin{aligned}
        \Delta^{\mathrm{AT}}_{u, i} = \epsilon \frac{\Gamma_{u, i}}{\|\Gamma_{u, i}\|}, \quad \text{where} \quad \Gamma_{u, i} = \frac{\partial \mathcal{L}((u, i)|\Theta + \Delta)}{\partial \Delta_{u, i}}.
    \end{aligned} 
\end{equation}

\section{Method}
\label{sec:method}

\begin{figure}
    \centering
    \subfigure[The number of times users are affected by attacks.]{
    \includegraphics[width=0.475\textwidth]{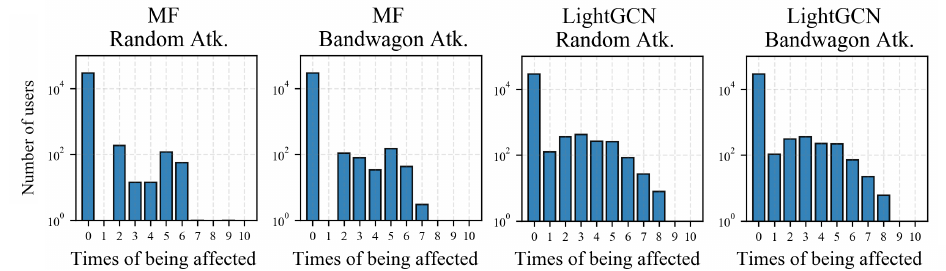}
    \label{fig:user_vul}
    }
    \subfigure[The number of the changes in users’ attack statuses for the users who have been affected.]{
    \includegraphics[width=0.475\textwidth]{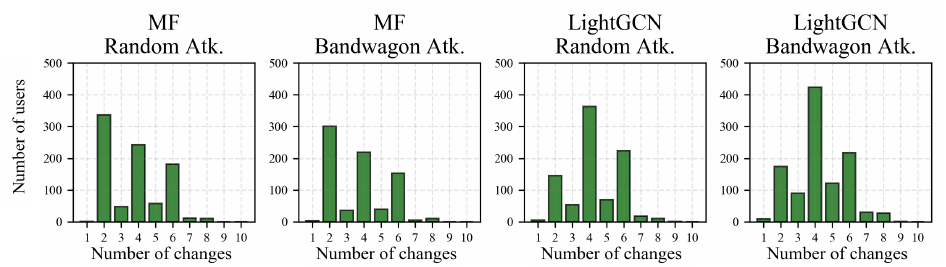}
    \label{fig:user_change}
    }
    \caption{User's vulnerability is fluctuant. (1) Few users consistently demonstrate vulnerability; (2) Most users who have been successfully attacked have multiple status changes.}
\end{figure}

To address the limitations of existing adversarial training methods, we propose user-adaptive magnitudes of perturbations, integrating large-magnitude perturbations for users vulnerable to attacks, thus offering effective protection. Simultaneously, we reduce the magnitude of adversarial perturbations for users deemed invulnerable, aiming to preserve the quality of their recommendations. This section delves into identifying these vulnerable users. Subsequently, we introduce the \textbf{Vulnerability-Aware Adversarial Training (VAT)} method, which tailors adversarial training to the specific vulnerabilities of users by applying perturbations of user-adaptive magnitudes. VAT aims to both provide effective protection and maintain recommendation performance by adapting to the nuanced needs of individual users.

\begin{figure*}
    \centering
    \subfigure[Proportion of users who are affected by attacks across different loss bins.]{
    \includegraphics[width=0.48\textwidth]{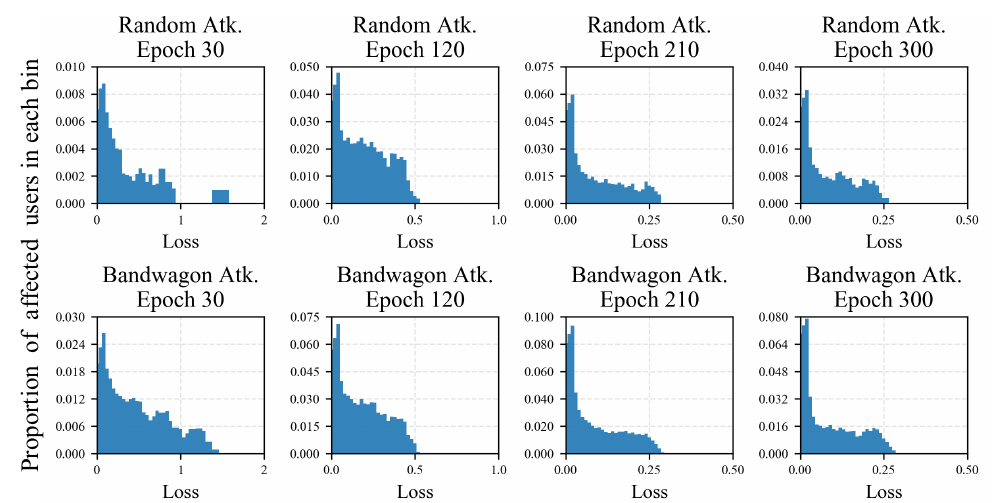}
    \label{fig:loss_dis}
    }
    \subfigure[Number of users who are affected by attacks across different loss bins.]{
    \includegraphics[width=0.48\textwidth]{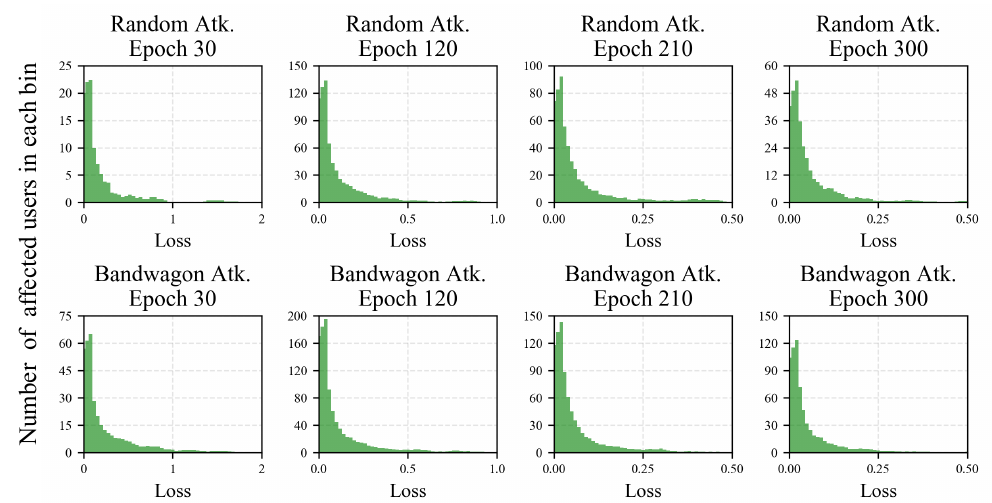}
    \label{fig:loss_num}
    }
    \caption{Users with lower losses are more likely to be affected by attacks in comparison to those with higher losses.}
    \label{fig:user_loss}
\end{figure*}

\subsection{User's Vulnerability Is Fluctuant}

To seek indicators to estimate user vulnerability, we initially examine whether this vulnerability is static, derived from user statistics, or fluctuant, evolving with the recommender system's training. Accordingly, we assess the frequency with which users are affected by attacks during the training process.

Using the Gowalla dataset~\cite{liang2016modeling, he2020lightgcn} as an example dataset, we implement both the Random Attack~\cite{lam2004shilling} and Bandwagon Attack~\cite{burke2005limited}, and evaluate their impact on Matrix Factorization (MF)~\cite{koren2009matrix} and LightGCN~\cite{he2020lightgcn} as victim models. For a detailed discussion of the experimental settings, please refer to Section~\ref{sec:exp_setup}. The recommender system is trained under these conditions across 300 epochs, during which we evaluate whether users are affected by attacks\footnote{A user is considered affected by attacks if any target item appears in the user's top 50 recommendation list~\cite{tang2020revisiting, huang2021data}.} every 30 epochs, resulting in a total of 10 evaluations.

Figure~\ref{fig:user_vul} presents the distribution of users' attack statuses (whether affected) over these 10 evaluations, illustrating that a predominant portion of users has never been affected (denoted by a horizontal coordinate of 0), while almost no users consistently demonstrate vulnerability (denoted by a horizontal coordinate of 10). Additionally, users affected by the attack have varying frequencies, with horizontal coordinates ranging from 1 to 9.

Moreover, Figure~\ref{fig:user_change}, which shows the changes in users' attack statuses\footnote{Changes in attack status are marked when there is a discrepancy between two successive evaluations, where the initial state is not being affected.} over time, indicates that a majority of those who have been affected undergo several status changes, with horizontal coordinates ranging from 2 to 8. This emphasizes the fluctuating nature of user vulnerability. These analyses, supported by Figure~\ref{fig:user_vul} and Figure~\ref{fig:user_change}, confirm that user vulnerability is indeed fluctuant during the training of the recommender system.

\subsection{Well-Fitted Users Are More Likely to Be Vulnerable}

Considering the fluctuant nature of user vulnerability during recommender system training, this section explores the relationship between a user's vulnerability to attacks and the training process of the recommender system.

\textbf{Hypothesis on User Vulnerability}.
Poisoning attacks manipulate the training of recommender systems by polluting the training data. These attacks establish deceptive correlations between users' historical interactions and the target items chosen by attackers. If the recommender system captures these deceptive correlations during the process of fitting user behavior, the user may be affected by attacks. This insight leads us to pose a critical question: \textit{Are well-fitted users in the current recommender system more likely to be affected by attacks}?

\textbf{Observation}.
To validate our hypothesis, we use user-specific loss as a measure of fit within the recommender system, regarding that users with lower loss values are better fitted by the system. We record each user's training loss alongside their attack status. Due to space constraints, we present the results of LightGCN on Gowalla in Figure~\ref{fig:user_loss}. Figure~\ref{fig:loss_dis} shows the percentage of users affected by attacks relative to the total user count within each loss bin\footnote{Only bins including more than 0.5\% of the total number of users are included to ensure visibility; bins below this threshold are excluded due to potential data instability.}. Meanwhile, Figure~\ref{fig:loss_num} displays the number of users affected by attacks across different loss bins. Our observations reveal that users with smaller losses have a higher probability of being affected compared to those with larger losses, indicating a general downward trend in the proportion and the number of users affected by attacks as loss increases. These observations empirically substantiate our hypothesis that \textit{well-fitted users in the current recommender system are more likely to be vulnerable, i.e., affected by attacks}.

\begin{figure}
    \centering
    \includegraphics[width=0.475\textwidth]{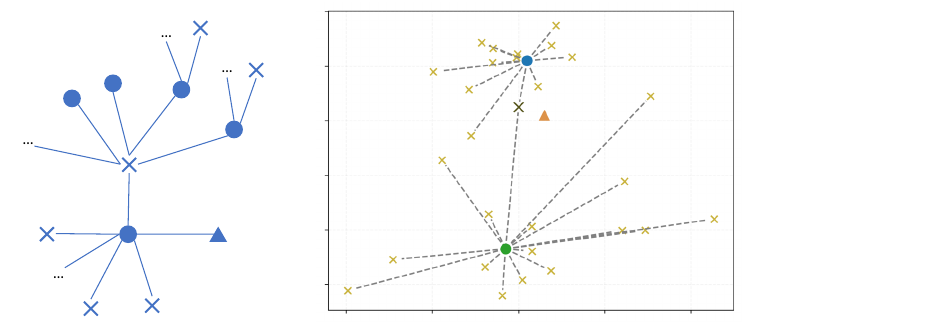}
    \caption{Embeddings of well-fitted user and under-fitted user. The cosine similarities in the original space for the two users to the target item are 0.7570 (well-fitted user) and 0.5309 (under-fitted user).}
    \label{fig:fit_case}
\end{figure}

\textbf{Analysis}.
Poisoning attacks exploit the system’s fitting capabilities by creating deceptive correlations between users' historical interactions and the attacker's chosen targets. These correlations may involve complex patterns, such as high-order connectivity with other items/users or similarities in consumption behavior, which the system is designed to capture and utilize. Figure~\ref{fig:fit_case} demonstrates a real example involving a well-fitted user (characterized by a small loss), an under-fitted user (characterized by a large loss), a fake user with a target item, and other normal items. By employing T-SNE~\cite{van2008visualizing} to project their embeddings into two dimensions, we observe that the well-fitted user precisely models the third-order link, thereby showing high similarity with the target item. Conversely, the under-fitted user fails to discern this pattern, remaining unaffected by the attack. In other words, users whose interactions are better fitted by the system more readily identify and utilize these deceptive correlations, increasing the likelihood of the attacker’s target items being recommended. Thus, increasing the magnitude of perturbations for these users is necessary to enhance their protection. This process intuitively explains why users who are well-fitted are more susceptible to the influence of poisoned data, making them more likely to be vulnerable to poisoning attacks.

\subsection{Vulnerability-Aware Adversarial Training}
\label{sec:VAT-train}

To enhance the robustness of users identified as vulnerable, we increase the magnitude of adversarial perturbations for these users to improve their ability to resist poisoning attacks, thus providing more effective protection. Recognizing that users with smaller losses are more likely to be vulnerable, we propose a vulnerability-aware function $g(\cdot)$ to quantify users’ vulnerability based on this indicator, which reflects such relatively small and large losses. To prevent excessively large perturbations over the training duration, we constrain $g: \mathbb{R} \rightarrow (0, 1)$. Formally, we define $g(\cdot)$ as follows:
\begin{equation}
    g\left(\mathcal{L}(u|\Theta)\right) = \sigma \left( \left( \frac{\mathcal{L}(u|\Theta) - \overline{\mathcal{L}(u|\Theta)}}{\overline{\mathcal{L}(u|\Theta)}} \right)^{-1}\right),
\end{equation}
where $\mathcal{L}(u|\Theta)$ denotes the loss associated with user $u$, $\overline{\mathcal{L}(u|\Theta)}$ is the mean loss across all users, and $\sigma(\cdot)$ is the Sigmoid function.

Given the recommender system's original loss function, $\mathcal{L}(\Theta)$, and referring to Equation~\ref{eq:at}, we integrate the vulnerability-aware function $g(\cdot)$ into $\Delta^{\mathrm{AT}}$. The loss function for VAT is expressed as follows:
\begin{equation}
\label{eq:vat}
    \begin{aligned}
        \mathcal{L}_{\mathrm{VAT}}(\Theta) =& \mathcal{L}(\Theta) + \lambda \mathcal{L}(\Theta + \Delta^{\mathrm{VAT}}), \\
        \mathrm{where} \quad \Delta^{\mathrm{VAT}} =& \arg \max_{\Delta,\, \Vert \Delta_{u, *} \Vert \leq \rho g(\mathcal{L}(u|\Theta))} \mathcal{L}(\Theta+\Delta),
    \end{aligned}  
\end{equation}
where $\lambda$ is the weight used in adversarial training, and $\rho$ determines the initial magnitude of perturbations. Specifically, for an interaction $(u, i)$, the perturbation of user-adaptive magnitude, $\Delta^{\mathrm{VAT}}_{u, i}$, is calculated as:
\begin{equation}
    \label{eq:vat_delta}
    \Delta^{\mathrm{VAT}}_{u, i} = \rho g\left(\mathcal{L}(u|\Theta)\right) \frac{\Gamma_{u, i}}{\|\Gamma_{u, i}\|}, \quad \text{where} \quad \Gamma_{u, i} = \frac{\partial \mathcal{L}((u, i)|\Theta + \Delta)}{\partial \Delta_{u, i}}.
\end{equation}
According to Equation~\ref{eq:vat_delta}, we apply such user-adaptive magnitudes of perturbations based on the user vulnerability, thereby providing an effective defense against poisoning attacks while maintaining the performance of the recommender system.

\subsection{Further Discussion}
It is important to note that although users with lower loss values are more vulnerable to attacks, the overall success ratio of these attacks remains low, leaving a part of low-loss users unaffected. Nonetheless, adversarial training at the parameter level also proves effective in cases where model parameters are overfitted to the data, as demonstrated in \cite{he2018adversarial}. For users with small losses, the introduction of large-magnitude perturbations can help correct the overfitting of parameters, thereby improving the quality of recommendations. With these dual benefits, VAT is capable of not only enhancing defenses for users vulnerable to attacks but also improving the generalization capabilities of the recommender system for users whose parameters are overfitted, as evident in Section~\ref{sec:over_fit}.

\section{EXPERIMENTS}
In this section, we conduct extensive experiments to answer the following research questions (\textbf{RQs}).
\begin{itemize}[leftmargin=*]
    \item \textbf{RQ1:} Can VAT defend against poisoning attacks?
    \item \textbf{RQ2:} How do hyper-parameters affect VAT?
    \item \textbf{RQ3:} Why does VAT outperform traditional adversarial training methods?
\end{itemize}

\subsection{Experimental Setup}
\label{sec:exp_setup}
\begin{table}[t]
  \centering
    \caption{Dataset statistics}
    \resizebox{0.47\textwidth}{!}{

\begin{tabular}{lrrrrr}
    \toprule
    \textbf{ DATASET } & \textbf{ \#Users } & \textbf{ \#Items } & \textbf{\#Ratings}  & \textbf{Avg.Inter.} & \textbf{Sparsity}\\
    \midrule
     Gowalla  & 29,858 & 40,981& 1,027,370 & 34.4 & 99.92\% \\
     Yelp2018  & 31,668 & 38,048 & 1,561,406 & 49.3 & 99.88\% \\ 
     MIND  & 141,920 & 36,214 & 20,693,122 & 145.8 & 99.60\% \\ 
    \bottomrule
    \end{tabular}
    }
  \label{tab:datasets}%
\end{table}%

\subsubsection{Datasets}
We employ three widely recognized datasets: the \textbf{Gowalla} check-in dataset~\cite{liang2016modeling}, the \textbf{Yelp2018} business dataset, and the \textbf{MIND} news recommendation dataset~\cite{wu2020mind}. The Gowalla and Yelp2018 datasets include all users, whereas for the MIND dataset, we sample a subset of users in alignment with~\cite{zhang2024lorec}. Consistent with \cite{he2020lightgcn, wang2019neural}, users and items with fewer than 10 interactions are excluded from our analysis. We allocate 80\% of each user's historical interactions to the training set and reserve the remainder for testing. Additionally, within the training set, 10\% of the interactions are randomly selected to form a validation set for hyperparameter tuning. Detailed statistics of the datasets are summarized in Table~\ref{tab:datasets}.

\subsubsection{Baselines for Defense}
We incorporate a variety of defense methods, including detection-based methods, adversarial training methods, and a denoise-based method. Specifically, we examine GraphRfi~\cite{zhang2020gcnbased} and LLM4Dec~\cite{zhang2024lorec} for detection-based methods; APR~\cite{he2018adversarial} and SharpCF~\cite{chen2023adversarial} for adversarial training methods; and StDenoise~\cite{tian2022learning, ye2023towards} for the denoise-based approach. 
\begin{itemize}[leftmargin=*]
    \item \textbf{GraphRfi}~\cite{zhang2020gcnbased}: Employs a combination of Graph Convolutional Networks and Neural Random Forests for identifying fraudsters.
    \item \textbf{LLM4Dec}~\cite{zhang2024lorec}: Utilizes an LLM-based framework for fraudster detection.
    \item \textbf{APR}~\cite{he2018adversarial}: Generates parameter perturbations and integrates these perturbations into training.
    \item \textbf{SharpCF}~\cite{chen2023adversarial}: Adopts a sharpness-aware minimization approach to refine the adversarial training process proposed by APR.
    \item \textbf{StDenoise}~\cite{tian2022learning, ye2023towards}: Applies a structural denoising technique that leverages the similarity between $\bm{p}_u$ and $\bm{q}_i$ for each $(u, i)$ pair, aiding in the removal of noise, as described in \cite{tian2022learning, ye2023towards}.
\end{itemize}
Note that LLM4Dec, which relies on item-side information, is exclusively evaluated on the MIND dataset. Additionally, we observe that SharpCF, initially proposed for the MF model, exhibits unstable training performance when applied to the LightGCN model or the MIND dataset. Consequently, we present SharpCF results solely for the MF model on the Gowalla and Yelp2018 datasets.

\begin{table*}[t]
    \centering
    \caption{Robustness against target items promotion}
    \resizebox{\textwidth}{!}{

\begin{tabular}{clcccccccc}
    \toprule
    \multicolumn{1}{c}{\multirow{2}{*}{\textbf{Dataset}}}& \multicolumn{1}{c}{\multirow{2}{*}{\textbf{Model}}} & \multicolumn{2}{c}{\textbf{Random Attack}(\%)} & \multicolumn{2}{c}{\textbf{Bandwagon Attack}(\%)} & \multicolumn{2}{c}{\textbf{DP Attack}(\%)} & \multicolumn{2}{c}{\textbf{Rev Attack}(\%)} \\ 
    \cmidrule(lr){3-4} \cmidrule(lr){5-6} \cmidrule(lr){7-8} \cmidrule(lr){9-10}
    & & \textbf{T-HR@50}$^1$ & \textbf{T-NDCG@50} & \textbf{T-HR@50} & \textbf{T-NDCG@50} & \textbf{T-HR@50} & \textbf{T-NDCG@50} & \textbf{T-HR@50} & \textbf{T-NDCG@50} \\
    \midrule
    \multirow{13}{1.2cm}{\centering \textbf{Gowalla}}
    &\textbf{MF}             &\underline{0.148 $\pm$ 0.030}&\underline{0.036 $\pm$ 0.008}&\underline{0.120 $\pm$ 0.027}&\underline{0.029 $\pm$ 0.007}&0.201 $\pm$ 0.020            &0.051 $\pm$ 0.005            &0.246 $\pm$ 0.097            &0.061 $\pm$ 0.027            \\
    &~~+\textbf{StDenoise}   &0.200 $\pm$ 0.049            &0.050 $\pm$ 0.012            &0.165 $\pm$ 0.034            &0.038 $\pm$ 0.008            &0.292 $\pm$ 0.034            &0.074 $\pm$ 0.010            &0.355 $\pm$ 0.126            &0.084 $\pm$ 0.030            \\
    &~~+\textbf{GraphRfi}    &0.159 $\pm$ 0.061            &0.042 $\pm$ 0.015            &0.154 $\pm$ 0.038            &0.036 $\pm$ 0.009            &0.174 $\pm$ 0.038            &0.043 $\pm$ 0.009            &\underline{0.206 $\pm$ 0.042}&\underline{0.050 $\pm$ 0.010}\\
    &~~+\textbf{APR}         &0.201 $\pm$ 0.091            &0.054 $\pm$ 0.026            &0.184 $\pm$ 0.067            &0.047 $\pm$ 0.015            &\underline{0.034 $\pm$ 0.021}&\underline{0.006 $\pm$ 0.004}&0.261 $\pm$ 0.063            &0.067 $\pm$ 0.018            \\
    &~~+\textbf{SharpCF}     &0.204 $\pm$ 0.037            &0.049 $\pm$ 0.010            &0.169 $\pm$ 0.031            &0.041 $\pm$ 0.008            &0.303 $\pm$ 0.024            &0.077 $\pm$ 0.006            &0.350 $\pm$ 0.111            &0.087 $\pm$ 0.031            \\
    \cmidrule{3-10}
    &~~+\textbf{VAT}         &\textbf{0.121 $\pm$ 0.028}  &\textbf{0.031 $\pm$ 0.009}  &\textbf{0.101 $\pm$ 0.038}  &\textbf{0.024 $\pm$ 0.008}  &\textbf{0.028 $\pm$ 0.007}  &\textbf{0.006 $\pm$ 0.001}  &\textbf{0.103 $\pm$ 0.048}  &\textbf{0.024 $\pm$ 0.011}  \\
    & \multicolumn{1}{c}{Gain$^2$}& +18.49\% $\uparrow$& +15.63\% $\uparrow$& +15.86\% $\uparrow$& +16.36\% $\uparrow$& +16.77\% $\uparrow$& +9.03\% $\uparrow$& +49.87\% $\uparrow$& +52.39\% $\uparrow$\\
    \cmidrule{2-10}
    &\textbf{LightGCN}       &0.234 $\pm$ 0.116            &0.056 $\pm$ 0.031            &0.639 $\pm$ 0.090            &0.153 $\pm$ 0.024            &0.231 $\pm$ 0.048            &0.048 $\pm$ 0.010            &0.718 $\pm$ 0.134            &0.149 $\pm$ 0.026            \\
    &~~+\textbf{StDenoise}   &0.118 $\pm$ 0.068            &0.029 $\pm$ 0.019            &0.334 $\pm$ 0.092            &\underline{0.079 $\pm$ 0.020}&0.585 $\pm$ 0.092            &0.120 $\pm$ 0.019            &1.304 $\pm$ 0.184            &0.259 $\pm$ 0.037            \\
    &~~+\textbf{GraphRfi}    &0.099 $\pm$ 0.023            &0.023 $\pm$ 0.006            &0.710 $\pm$ 0.250            &0.161 $\pm$ 0.052            &0.228 $\pm$ 0.048            &0.046 $\pm$ 0.010            &\underline{0.564 $\pm$ 0.067}&\underline{0.115 $\pm$ 0.013}\\
    &~~+\textbf{APR}         &\underline{0.090 $\pm$ 0.053}&\underline{0.022 $\pm$ 0.015}&\underline{0.332 $\pm$ 0.050}&0.079 $\pm$ 0.012            &\underline{0.190 $\pm$ 0.037}&\underline{0.039 $\pm$ 0.008}&0.655 $\pm$ 0.141            &0.132 $\pm$ 0.027            \\
    \cmidrule{3-10}
    &~~+\textbf{VAT}         &\textbf{0.089 $\pm$ 0.054}  &\textbf{0.021 $\pm$ 0.014}  &\textbf{0.259 $\pm$ 0.047}  &\textbf{0.063 $\pm$ 0.012}  &\textbf{0.141 $\pm$ 0.034}  &\textbf{0.028 $\pm$ 0.007}  &\textbf{0.456 $\pm$ 0.093}  &\textbf{0.094 $\pm$ 0.018}  \\
    & \multicolumn{1}{c}{Gain}& +0.22\% $\uparrow$& +0.55\% $\uparrow$& +22.01\% $\uparrow$& +20.77\% $\uparrow$& +25.86\% $\uparrow$& +28.32\% $\uparrow$& +19.17\% $\uparrow$& +18.29\% $\uparrow$\\
    \midrule
    \multirow{13}{1.2cm}{\centering \textbf{Yelp2018}}
    &\textbf{MF}             &0.035 $\pm$ 0.007            &0.010 $\pm$ 0.002            &0.073 $\pm$ 0.032            &0.020 $\pm$ 0.009            &0.223 $\pm$ 0.040            &0.049 $\pm$ 0.009            &0.153 $\pm$ 0.025            &0.040 $\pm$ 0.006            \\
    &~~+\textbf{StDenoise}   &0.015 $\pm$ 0.038            &0.007 $\pm$ 0.010            &0.181 $\pm$ 0.046            &0.043 $\pm$ 0.011            &0.376 $\pm$ 0.198            &0.077 $\pm$ 0.039            &0.331 $\pm$ 0.145            &0.075 $\pm$ 0.031            \\
    &~~+\textbf{GraphRfi}    &0.032 $\pm$ 0.009            &0.009 $\pm$ 0.003            &0.058 $\pm$ 0.014            &0.015 $\pm$ 0.003            &0.200 $\pm$ 0.041            &0.043 $\pm$ 0.010            &0.129 $\pm$ 0.027            &0.031 $\pm$ 0.007            \\
    &~~+\textbf{APR}         &\underline{0.012 $\pm$ 0.007}&\underline{0.004 $\pm$ 0.002}&\underline{0.057 $\pm$ 0.047}&\underline{0.013 $\pm$ 0.011}&\underline{0.185 $\pm$ 0.038}&\underline{0.040 $\pm$ 0.009}&\underline{0.098 $\pm$ 0.048}&\underline{0.022 $\pm$ 0.011}\\
    &~~+\textbf{SharpCF}     &0.034 $\pm$ 0.007            &0.010 $\pm$ 0.002            &0.072 $\pm$ 0.029            &0.019 $\pm$ 0.008            &0.226 $\pm$ 0.041            &0.050 $\pm$ 0.010            &0.152 $\pm$ 0.025            &0.040 $\pm$ 0.006            \\
    \cmidrule{3-10}
    &~~+\textbf{VAT}         &\textbf{0.010 $\pm$ 0.006}  &\textbf{0.003 $\pm$ 0.002}  &\textbf{0.040 $\pm$ 0.031}  &\textbf{0.010 $\pm$ 0.007}  &\textbf{0.142 $\pm$ 0.038}  &\textbf{0.028 $\pm$ 0.007}  &\textbf{0.090 $\pm$ 0.049}  &\textbf{0.020 $\pm$ 0.010}  \\
    & \multicolumn{1}{c}{Gain}& +14.11\% $\uparrow$& +19.16\% $\uparrow$& +30.70\% $\uparrow$& +25.30\% $\uparrow$& +23.32\% $\uparrow$& +28.80\% $\uparrow$& +8.43\% $\uparrow$& +8.50\% $\uparrow$\\
    \cmidrule{2-10}
    &\textbf{LightGCN}       &0.381 $\pm$ 0.064            &0.116 $\pm$ 0.022            &1.286 $\pm$ 0.351            &0.299 $\pm$ 0.083            &0.451 $\pm$ 0.040            &0.098 $\pm$ 0.008            &1.761 $\pm$ 0.368            &0.402 $\pm$ 0.091            \\
    &~~+\textbf{StDenoise}   &\textbf{0.058 $\pm$ 0.017}&\textbf{0.018 $\pm$ 0.008}&1.609 $\pm$ 0.381            &0.346 $\pm$ 0.091            &3.939 $\pm$ 0.417            &0.814 $\pm$ 0.094            &5.965 $\pm$ 0.375            &1.472 $\pm$ 0.125            \\
    &~~+\textbf{GraphRfi}    &0.434 $\pm$ 0.074            &0.127 $\pm$ 0.023            &\underline{0.958 $\pm$ 0.199}&\underline{0.200 $\pm$ 0.042}&0.581 $\pm$ 0.049            &0.119 $\pm$ 0.011            &1.597 $\pm$ 0.087            &0.344 $\pm$ 0.016            \\
    &~~+\textbf{APR}         &0.291 $\pm$ 0.050            &0.090 $\pm$ 0.018            &1.052 $\pm$ 0.278            &0.242 $\pm$ 0.065            &\underline{0.370 $\pm$ 0.034}&\underline{0.078 $\pm$ 0.007}&\underline{1.139 $\pm$ 0.179}&\underline{0.249 $\pm$ 0.041}\\
    \cmidrule{3-10}
    &~~+\textbf{VAT}         &\underline{0.082 $\pm$ 0.020}  &\underline{0.024 $\pm$ 0.006}  &\textbf{0.694 $\pm$ 0.181}  &\textbf{0.156 $\pm$ 0.041}  &\textbf{0.365 $\pm$ 0.037}  &\textbf{0.076 $\pm$ 0.008}  &\textbf{0.927 $\pm$ 0.135}  &\textbf{0.196 $\pm$ 0.029}  \\
    & \multicolumn{1}{c}{Gain}& - & - & +27.56\% $\uparrow$& +22.13\% $\uparrow$& +1.50\% $\uparrow$& +1.88\% $\uparrow$& +18.61\% $\uparrow$& +21.25\% $\uparrow$\\
    \midrule
    \multirow{13}{1.2cm}{\centering \textbf{MIND}}
    &\textbf{MF}             &0.032 $\pm$ 0.007            &0.010 $\pm$ 0.002            &0.169 $\pm$ 0.017            &0.055 $\pm$ 0.005            &0.023 $\pm$ 0.013            &0.005 $\pm$ 0.003            & { \footnotesize OOM}$^3$ & { \footnotesize OOM}\\
    &~~+\textbf{StDenoise}   
    &0.036 $\pm$ 0.006            &0.013 $\pm$ 0.004            &\underline{0.040 $\pm$ 0.006}           &\underline{0.020 $\pm$ 0.004}           &0.010 $\pm$ 0.003            &0.002 $\pm$ 0.001 & { \footnotesize OOM} & { \footnotesize OOM}\\
    &~~+\textbf{GraphRfi}    &0.031 $\pm$ 0.006            &0.010 $\pm$ 0.002            &0.189 $\pm$ 0.015            &0.059 $\pm$ 0.005            &0.020 $\pm$ 0.009            &0.004 $\pm$ 0.002            & { \footnotesize OOM} & { \footnotesize OOM}\\
    &~~+\textbf{LLM4Dec}     &\textbf{0.020 $\pm$ 0.001}&\textbf{0.004 $\pm$ 0.000}&0.083 $\pm$ 0.009            &0.025 $\pm$ 0.003            &0.019 $\pm$ 0.010            &0.004 $\pm$ 0.002            & { \footnotesize OOM} & { \footnotesize OOM}\\
    &~~+\textbf{APR}         &0.083 $\pm$ 0.013            &0.035 $\pm$ 0.006            &0.068 $\pm$ 0.005            &0.023 $\pm$ 0.002            &\underline{0.008 $\pm$ 0.007}&\underline{0.002 $\pm$ 0.001}& { \footnotesize OOM} & { \footnotesize OOM}\\
    \cmidrule{3-10}
    &~~+\textbf{VAT}         &\underline{0.026 $\pm$ 0.006}  &\underline{0.011 $\pm$ 0.003}  &\textbf{0.032 $\pm$ 0.004}  &\textbf{0.011 $\pm$ 0.001}  &\textbf{0.002 $\pm$ 0.002}  &\textbf{0.000 $\pm$ 0.000}  & { \footnotesize OOM} & { \footnotesize OOM}\\
    & \multicolumn{1}{c}{Gain}& - & - & +20.15\% $\uparrow$& +45.40\% $\uparrow$& +75.36\% $\uparrow$& +77.27\% $\uparrow$& - & -\\
    \cmidrule{2-10}
    &\textbf{LightGCN}       &0.056 $\pm$ 0.008            &0.015 $\pm$ 0.002            &0.149 $\pm$ 0.016            &0.038 $\pm$ 0.004            &0.006 $\pm$ 0.002            &0.001 $\pm$ 0.000            & { \footnotesize OOM} & { \footnotesize OOM}\\
    &~~+\textbf{StDenoise}   &0.052 $\pm$ 0.026            &0.014 $\pm$ 0.020            &0.164 $\pm$ 0.017            &0.040 $\pm$ 0.004            &0.007 $\pm$ 0.002            &0.001 $\pm$ 0.001            & { \footnotesize OOM} & { \footnotesize OOM}\\
    &~~+\textbf{GraphRfi}    &0.045 $\pm$ 0.004            &\underline{0.012 $\pm$ 0.001}&0.093 $\pm$ 0.007            &\underline{0.022 $\pm$ 0.001}&0.008 $\pm$ 0.001            &0.002 $\pm$ 0.000            & { \footnotesize OOM} & { \footnotesize OOM}\\
    &~~+\textbf{LLM4Dec}     &\underline{0.039 $\pm$ 0.017}&0.013 $\pm$ 0.006            &0.104 $\pm$ 0.009            &0.027 $\pm$ 0.003            &0.006 $\pm$ 0.002            &0.001 $\pm$ 0.001            & { \footnotesize OOM} & { \footnotesize OOM}\\
    &~~+\textbf{APR}         &0.053 $\pm$ 0.007            &0.016 $\pm$ 0.002            &\underline{0.091 $\pm$ 0.007}&0.022 $\pm$ 0.002            &\underline{0.006 $\pm$ 0.001}&\underline{0.001 $\pm$ 0.000}& { \footnotesize OOM} & { \footnotesize OOM}\\
    \cmidrule{3-10}
    &~~+\textbf{VAT}         &\textbf{0.032 $\pm$ 0.005}  &\textbf{0.010 $\pm$ 0.001}  &\textbf{0.065 $\pm$ 0.005}  &\textbf{0.016 $\pm$ 0.001}  &\textbf{0.003 $\pm$ 0.001}  &\textbf{0.001 $\pm$ 0.000}  & { \footnotesize OOM} & { \footnotesize OOM}\\
    & \multicolumn{1}{c}{Gain}& +17.20\% $\uparrow$& +18.64\% $\uparrow$& +28.50\% $\uparrow$& +26.56\% $\uparrow$& +40.00\% $\uparrow$& +39.66\% $\uparrow$& - & -\\
    \bottomrule
\end{tabular}
   
    }
\label{tab:attack_per}%
{ \footnotesize
\raggedright
1. Target Item Hit Ratio (Equation~\ref{eq:tar}); T-HR@50 and T-NDCG@50 of all target items on clean datasets are 0.000. \\
~~2. The relative percentage increase of VAT's metrics to the best value of other baselines' metrics, i.e., $\left(\min\left(\mathrm{T}\text{-}\mathrm{HR}_\mathrm{Beslines}\right) - \mathrm{T}\text{-}\mathrm{HR}_\mathrm{VAT} \right)/ \min(\mathrm{T}\text{-}\mathrm{HR}_\mathrm{Beslines})$. Notably, only \textbf{three decimal places} are presented due to space limitations, though the actual ranking and calculations utilize the \textbf{full precision} of the data. \\
~~3. The Rev attack method could not be executed on the dataset due to memory constraints, resulting in an out-of-memory error.\\
}
\end{table*}

\begin{figure}
    \centering
    \includegraphics[width=0.475\textwidth]{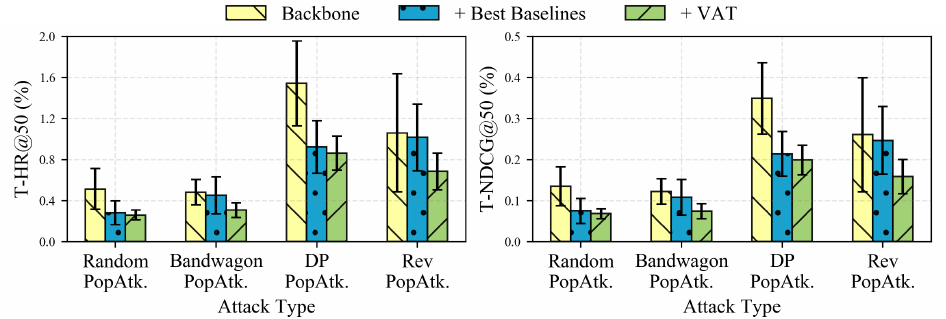}
    \caption{Robustness against popular items promotion.}
    \label{fig:pop_attack}
\end{figure}

\begin{table*}[t]
    \centering
    \caption{Recommendation performance}
    \resizebox{\textwidth}{!}{

\begin{tabular}{lcccccccccc}
    \toprule
    \multicolumn{1}{c}{\multirow{1}{*}{\textbf{Model}}} & \multicolumn{2}{c}{\textbf{Clean} (\%)} & \multicolumn{2}{c}{\textbf{Random Attack} (\%)} & \multicolumn{2}{c}{\textbf{Bandwagon Attack} (\%)} & \multicolumn{2}{c}{\textbf{DP Attack} (\%)} & \multicolumn{2}{c}{\textbf{Rev Attack} (\%)}  \\ 
    \cmidrule(lr){2-3} \cmidrule(lr){4-5} \cmidrule(lr){6-7} \cmidrule(lr){8-9}  \cmidrule(lr){10-11}
    \multicolumn{1}{c}{\multirow{1}{*}{(Dataset)}} & \textbf{HR@20} & \textbf{NDCG@20} & \textbf{HR@20} & \textbf{NDCG@20} & \textbf{HR@20} & \textbf{NDCG@20} & \textbf{HR@20} & \textbf{NDCG@20} & \textbf{HR@20} & \textbf{NDCG@20}  \\
    \midrule
    \textbf{MF}~~(Gowalla)            &11.352 $\pm$ 0.091            &7.158 $\pm$ 0.035            &11.306 $\pm$ 0.077            &7.196 $\pm$ 0.061            &11.238 $\pm$ 0.077            &7.106 $\pm$ 0.042            &10.722 $\pm$ 0.109            &8.170 $\pm$ 0.076            &10.698 $\pm$ 0.090            &8.188 $\pm$ 0.044            \\
    ~~+\textbf{StDenoise}    &10.484 $\pm$ 0.096            &8.074 $\pm$ 0.103            &10.456 $\pm$ 0.089            &8.074 $\pm$ 0.067            &10.412 $\pm$ 0.058            &8.038 $\pm$ 0.023            &10.532 $\pm$ 0.130            &8.120 $\pm$ 0.089            &10.568 $\pm$ 0.047            &8.186 $\pm$ 0.038            \\
    ~~+\textbf{GraphRfi}     &10.434 $\pm$ 0.065            &7.968 $\pm$ 0.026            &10.344 $\pm$ 0.080            &7.886 $\pm$ 0.057            &10.304 $\pm$ 0.059            &7.846 $\pm$ 0.061            &10.400 $\pm$ 0.115            &7.942 $\pm$ 0.079            &10.496 $\pm$ 0.093            &8.010 $\pm$ 0.069            \\
    ~~+\textbf{APR}          &13.058 $\pm$ 0.063            &\underline{10.646 $\pm$ 0.058}&12.934 $\pm$ 0.044            &\underline{10.520 $\pm$ 0.013}&12.902 $\pm$ 0.065            &\underline{10.500 $\pm$ 0.030}&12.946 $\pm$ 0.056            &\underline{10.586 $\pm$ 0.060}&13.128 $\pm$ 0.052            &\underline{10.720 $\pm$ 0.065}\\
    ~~+\textbf{SharpCF}      &\underline{13.203 $\pm$ 0.074}&10.020 $\pm$ 0.090            &\underline{13.188 $\pm$ 0.077}&10.028 $\pm$ 0.069            &\underline{13.025 $\pm$ 0.060}&9.890 $\pm$ 0.050            &\underline{13.270 $\pm$ 0.138}&10.082 $\pm$ 0.098            &\underline{13.215 $\pm$ 0.087}&10.095 $\pm$ 0.044            \\
    \cmidrule{2-11}
    ~~+\textbf{VAT}          &\textbf{13.424 $\pm$ 0.041}  &\textbf{10.864 $\pm$ 0.047}  &\textbf{13.292 $\pm$ 0.016}  &\textbf{10.764 $\pm$ 0.012}  &\textbf{13.286 $\pm$ 0.029}  &\textbf{10.740 $\pm$ 0.018}  &\textbf{13.396 $\pm$ 0.045}  &\textbf{10.860 $\pm$ 0.036}  &\textbf{13.540 $\pm$ 0.087}  &\textbf{10.980 $\pm$ 0.059}  \\
    \multicolumn{1}{c}{Gain}& +1.68\% $\uparrow$& +2.05\% $\uparrow$& +0.79\% $\uparrow$& +2.32\% $\uparrow$& +2.00\% $\uparrow$& +2.29\% $\uparrow$& +0.95\% $\uparrow$& +2.59\% $\uparrow$& +2.46\% $\uparrow$& +2.43\% $\uparrow$\\
    \multicolumn{1}{c}{Gain w.r.t. MF}& +18.25\% $\uparrow$& +51.77\% $\uparrow$& +17.57\% $\uparrow$& +49.58\% $\uparrow$& +18.22\% $\uparrow$& +51.14\% $\uparrow$& +24.94\% $\uparrow$& +32.93\% $\uparrow$& +26.57\% $\uparrow$& +34.10\% $\uparrow$\\
    \midrule
    \textbf{MF}~~(Yelp2018)             &3.762 $\pm$ 0.034            &2.974 $\pm$ 0.039            &3.730 $\pm$ 0.017            &2.934 $\pm$ 0.010            &3.744 $\pm$ 0.040            &2.948 $\pm$ 0.029            &3.866 $\pm$ 0.038            &3.028 $\pm$ 0.033            &3.812 $\pm$ 0.044            &3.028 $\pm$ 0.041            \\
    ~~+\textbf{StDenoise}    &3.410 $\pm$ 0.085            &2.612 $\pm$ 0.092            &3.288 $\pm$ 0.040            &2.504 $\pm$ 0.026            &3.322 $\pm$ 0.057            &2.522 $\pm$ 0.047            &3.384 $\pm$ 0.062            &2.578 $\pm$ 0.063            &3.380 $\pm$ 0.104            &2.586 $\pm$ 0.102            \\
    ~~+\textbf{GraphRfi}     &3.726 $\pm$ 0.051            &2.942 $\pm$ 0.034            &3.664 $\pm$ 0.038            &2.902 $\pm$ 0.033            &3.640 $\pm$ 0.054            &2.882 $\pm$ 0.029            &3.762 $\pm$ 0.056            &2.932 $\pm$ 0.049            &3.718 $\pm$ 0.053            &2.950 $\pm$ 0.042            \\
    ~~+\textbf{APR}          &\underline{4.094 $\pm$ 0.022}&\underline{3.202 $\pm$ 0.017}&\underline{4.036 $\pm$ 0.019}&\underline{3.160 $\pm$ 0.018}&\underline{4.080 $\pm$ 0.028}&\underline{3.194 $\pm$ 0.026}&4.012 $\pm$ 0.059            &3.152 $\pm$ 0.043            &\underline{4.061 $\pm$ 0.029}&\underline{3.205 $\pm$ 0.024}\\
    ~~+\textbf{SharpCF}      &3.933 $\pm$ 0.038            &3.108 $\pm$ 0.045            &3.883 $\pm$ 0.015            &3.058 $\pm$ 0.016            &3.910 $\pm$ 0.051            &3.079 $\pm$ 0.027            &\underline{4.034 $\pm$ 0.034}&\underline{3.161 $\pm$ 0.037}&3.971 $\pm$ 0.052            &3.156 $\pm$ 0.047            \\
    \cmidrule{2-11}
    ~~+\textbf{VAT}          &\textbf{4.112 $\pm$ 0.023}  &\textbf{3.234 $\pm$ 0.022}  &\textbf{4.074 $\pm$ 0.016}  &\textbf{3.206 $\pm$ 0.014}  &\textbf{4.130 $\pm$ 0.035}  &\textbf{3.246 $\pm$ 0.030}  &\textbf{4.096 $\pm$ 0.044}  &\textbf{3.202 $\pm$ 0.041}  &\textbf{4.218 $\pm$ 0.027}  &\textbf{3.326 $\pm$ 0.024}  \\
    \multicolumn{1}{c}{Gain}& +0.44\% $\uparrow$& +1.00\% $\uparrow$& +0.94\% $\uparrow$& +1.46\% $\uparrow$& +1.23\% $\uparrow$& +1.63\% $\uparrow$& +1.53\% $\uparrow$& +1.31\% $\uparrow$& +3.86\% $\uparrow$& +3.79\% $\uparrow$\\
    \multicolumn{1}{c}{Gain w.r.t. MF}& +9.30\% $\uparrow$& +8.74\% $\uparrow$& +9.22\% $\uparrow$& +9.27\% $\uparrow$& +10.31\% $\uparrow$& +10.11\% $\uparrow$& +5.95\% $\uparrow$& +5.75\% $\uparrow$& +10.65\% $\uparrow$& +9.84\% $\uparrow$\\
    \midrule
    \textbf{MF}~~(MIND)             &1.204 $\pm$ 0.014            &0.676 $\pm$ 0.005            &1.190 $\pm$ 0.011            &0.670 $\pm$ 0.006            &1.192 $\pm$ 0.016            &0.676 $\pm$ 0.005            &1.204 $\pm$ 0.005            &0.688 $\pm$ 0.007            & { \footnotesize OOM} & { \footnotesize OOM}\\
    ~~+\textbf{StDenoise}    &1.126 $\pm$ 0.014            &0.630 $\pm$ 0.006            &1.120 $\pm$ 0.006            &0.626 $\pm$ 0.005            &1.116 $\pm$ 0.008            &0.632 $\pm$ 0.004            &1.130 $\pm$ 0.006            &0.642 $\pm$ 0.007            & { \footnotesize OOM} & { \footnotesize OOM}\\
    ~~+\textbf{GraphRfi}     &1.198 $\pm$ 0.015            &0.666 $\pm$ 0.005            &1.188 $\pm$ 0.010            &0.666 $\pm$ 0.005            &1.194 $\pm$ 0.010            &0.668 $\pm$ 0.007            &1.204 $\pm$ 0.019            &0.674 $\pm$ 0.008            & { \footnotesize OOM} & { \footnotesize OOM}\\
    ~~+\textbf{LLM4Dec}      &1.200 $\pm$ 0.011            &0.676 $\pm$ 0.005            &1.190 $\pm$ 0.011            &0.670 $\pm$ 0.006            &1.194 $\pm$ 0.015            &0.676 $\pm$ 0.005            &1.194 $\pm$ 0.005            &0.682 $\pm$ 0.004            & { \footnotesize OOM} & { \footnotesize OOM}\\
    ~~+\textbf{APR}          &\underline{1.218 $\pm$ 0.010}&\underline{0.682 $\pm$ 0.007}&\underline{1.262 $\pm$ 0.016}&\underline{0.712 $\pm$ 0.007}&\underline{1.212 $\pm$ 0.008}&\underline{0.686 $\pm$ 0.004}&\underline{1.214 $\pm$ 0.010}&\underline{0.696 $\pm$ 0.008}& { \footnotesize OOM} & { \footnotesize OOM}\\
    \cmidrule{2-11}
    ~~+\textbf{VAT}          &\textbf{1.264 $\pm$ 0.012}  &\textbf{0.710 $\pm$ 0.000}  &\textbf{1.264 $\pm$ 0.014}  &\textbf{0.714 $\pm$ 0.005}  &\textbf{1.266 $\pm$ 0.008}  &\textbf{0.714 $\pm$ 0.005}  &\textbf{1.260 $\pm$ 0.013}  &\textbf{0.718 $\pm$ 0.010}  & { \footnotesize OOM} & { \footnotesize OOM}\\
    \multicolumn{1}{c}{Gain}& +3.78\% $\uparrow$& +4.11\% $\uparrow$& +0.16\% $\uparrow$& +0.28\% $\uparrow$& +4.44\% $\uparrow$& +4.10\% $\uparrow$& +3.79\% $\uparrow$& +3.16\% $\uparrow$& - & -\\
    \multicolumn{1}{c}{Gain w.r.t. MF}& +4.98\% $\uparrow$& +5.03\% $\uparrow$& +6.22\% $\uparrow$& +6.57\% $\uparrow$& +6.21\% $\uparrow$& +5.62\% $\uparrow$& +4.65\% $\uparrow$& +4.36\% $\uparrow$& - & -\\
    \bottomrule

\end{tabular}
   
    }
\label{tab:attack_rc}%
\end{table*}

\subsubsection{Attack Methods}
In our analysis, we explore both heuristic (Random Attack~\cite{lam2004shilling}, Bandwagon Attack~\cite{burke2005limited}) and optimization-based (Rev Attack~\cite{tang2020revisiting}, DP Attack~\cite{huang2021data}) attack methods within a black-box context, where the attacker does not have access to the internal architecture or parameters of the target model.

\subsubsection{Evaluation Metrics}
We adopt standard metrics widely accepted in the field. The primary metrics for assessing recommendation performance are the top-$k$ metrics: Hit Ratio at $k$ ($\mathrm{HR}@k$) and Normalized Discounted Cumulative Gain at $k$ ($\mathrm{NDCG}@k$), as documented in \cite{zhang2023robust, he2020lightgcn, wang2019neural}.
To quantify the success ratio of attacks, we utilize metrics tailored to measuring the performance of target items promotion within the top-$k$ recommendations, denoted as $\mathrm{T}\text{-}\mathrm{HR}@k$ and $\mathrm{T}\text{-}\mathrm{NDCG}@k$~\cite{tang2020revisiting, huang2021data, zhang2024lorec}:
\begin{equation}
    \label{eq:tar}
    \mathrm{T}\text{-}\mathrm{HR}@k = \frac{1}{|\mathcal{T}|} \sum_{\mathit{tar} \in \mathcal{T}} \frac{ \sum_{u \in \mathcal{U} - \mathcal{U}_{\mathit{tar}}} \mathbb{I}\left(\mathit{tar} \in L_{u, {1:k}}\right)}{|\mathcal{U} - \mathcal{U}_{\mathit{tar}}|},
\end{equation}
where $\mathcal{T}$ is the set of target items, $\mathcal{U}_{\mathit{tar}}$ denotes the set of genuine users interacted with target items $\mathit{tar}$, $L_{u, {1:k}}$ represents the top-$k$ list of recommendations for user $u$, and $\mathbb{I}(\cdot)$ is the indicator function that returns 1 if the condition is true. $\mathrm{T}\text{-}\mathrm{NDCG}@k$ mirrors $\mathrm{T}\text{-}\mathrm{HR}@k$, serving as the target item-specific version of $\mathrm{NDCG}@k$.

\subsubsection{Implementation Details}
In our study, we employ two common backbone recommendation models, MF~\cite{koren2009matrix} and LightGCN~\cite{he2020lightgcn}. To quantify the success ratio of attacks, we select $k=50$ as the evaluation metric following~\cite{huang2021data, tang2020revisiting, wu2021fight}, while for assessing recommendation performance, we utilize $k=20$ following~\cite{he2020lightgcn, wang2019neural}. The configuration of both the defense methods and the recommendation models involves selecting a learning rate from \{0.1, 0.01, $\dots$, $1 \times 10^{-5}$\}, and a weight decay from \{0, 0.1, $\dots$, $1 \times 10^{-5}$\}. The implementation of GraphRfi follows its paper. For the detection-based methods, we employ the Random Attack to generate supervised fraudster data. The magnitude parameter of adversarial perturbations in both APR and VAT is determined from a range of \{0.1, 0.2, $\dots$, 1.0\}. In terms of attack methods, we set the attack budget to $1\%$ and target five items. The hyperparameters align with those detailed in their original publications. Our implementation code is accessible via the provided link\footnote{\url{https://github.com/Kaike-Zhang/VAT}}.

\subsection{Performance Comparison~(RQ1)}
In this section, we answer \textbf{RQ1}. We focus on two key aspects: the robustness against poisoning attacks and the recommendation performance of our proposed VAT.

\subsubsection{Robustness Against Poisoning Attacks}
We evaluate the effectiveness of VAT in defending against poisoning attacks by analyzing the attack success ratio. Our experiments focus on items with extremely low popularity, indicated by $\mathrm{T}\text{-}\mathrm{HR}@50$ and $\mathrm{T}\text{-}\mathrm{NDCG}@50$ scores of 0.0 in the absence of any attack. Note: Lower scores of $\mathrm{T}\text{-}\mathrm{HR}@50$ and $\mathrm{T}\text{-}\mathrm{NDCG}@50$ signify stronger defense capabilities.

\textbf{Item Promotion Attack - unpopular Items}: The results in Table~\ref{tab:attack_per} reveal that the purely denoise-based defense strategy is mostly effective against random attacks, attributable to the random selection of items and the simplified task of filtering out these fake users' interactions. However, when faced with other types of attacks, denoise-based defenses might even increase the attack's success ratio. Detection-based methods, such as GraphRfi and LLM4Dec, demonstrate robust defense capabilities against attacks that align with their training data (notably, random attacks). However, the effectiveness of GraphRfi significantly diminishes against other types of attacks. In contrast, adversarial training methods, which do not rely on prior knowledge, consistently show stable defense against various attacks. Among them, VAT significantly outperforms traditional adversarial training methods like APR and SharpCF. VAT reduces the success ratio of attacks, decreasing an average of $\mathrm{T}\text{-}\mathrm{HR}@50$ and $\mathrm{T}\text{-}\mathrm{NDCG}@50$ by 21.53\% and 22.54\%, respectively, compared to the top baseline results. These findings underscore VAT's superior defense mechanism.

\textbf{Item Promotion Attack - Popular Items}: Furthermore, we assess VAT's defense capabilities against attacks targeting popular items on Gowalla. According to Figure~\ref{fig:pop_attack}, VAT exhibits strong defensibility, outperforming the best baseline even when attacks specifically promote popular items.

\begin{table}[t]
    \centering
    \caption{Robustness and Performance against Adaptive Attack}
    \resizebox{0.47\textwidth}{!}{

\begin{tabular}{lcccc}
    \toprule
    \textbf{Model} & \textbf{T-HR@50} (\%) & \textbf{T-NDCG@20} (\%) & \textbf{HR@20} (\%) & \textbf{NDCG@20} (\%) \\
    \midrule
    \textbf{MF} & 0.201 $\pm$ 0.020 & 0.051 $\pm$ 0.005 & 10.722 $\pm$ 0.109 & 8.170 $\pm$ 0.076 \\
    ~~+\textbf{TopBaseline} & 0.049 $\pm$ 0.024 & 0.012 $\pm$ 0.005 & 12.952 $\pm$ 0.082 & 10.630 $\pm$ 0.066 \\
    ~~+\textbf{VAT} & 0.033 $\pm$ 0.004 & 0.008 $\pm$ 0.001 & 13.461 $\pm$ 0.045 & 10.973 $\pm$ 0.040 \\
    \textbf{Gain} & +32.72\% $\uparrow$ & +34.45\% $\uparrow$ & +3.9\% $\uparrow$ & +3.2\% $\uparrow$ \\
    \bottomrule
\end{tabular}
   
    }
\label{tab:attack_ap}%
\end{table}

\textbf{Adaptive Item Promotion Attack}: Additionally, we evaluate the effectiveness of defenses against attacks generated by adaptive DP Attacks (note that the Rev Attack cannot be adaptive due to its close dependency on the loss function), as shown in Table~\ref{tab:attack_ap}. Our results indicate that VAT performs better in adaptive DP Attacks compared to non-adaptive ones, highlighting VAT's superior defense capability.

\subsubsection{Recommendation Performance}
In our assessment of the efficacy of various defense methods on recommendation performance, as depicted in Table~\ref{tab:attack_rc}, we observe a notable improvement in recommendation quality with the use of adversarial training methods. This observation aligns with findings from previous studies~\cite{he2018adversarial, chen2023adversarial}, which indicate that adversarial training can significantly enhance the performance of recommender systems. Among the evaluated methods, VAT stands out by achieving the most impressive outcomes in enhancing recommendation performance, surpassing other baseline approaches. This indicates that the user-adaptive magnitude of perturbations, while resisting attacks, can also positively impact recommendation performance.

\subsection{Hyper-Parameters Analysis~(RQ2)}
\begin{figure*}
    \centering
    \includegraphics[width=6.4in]{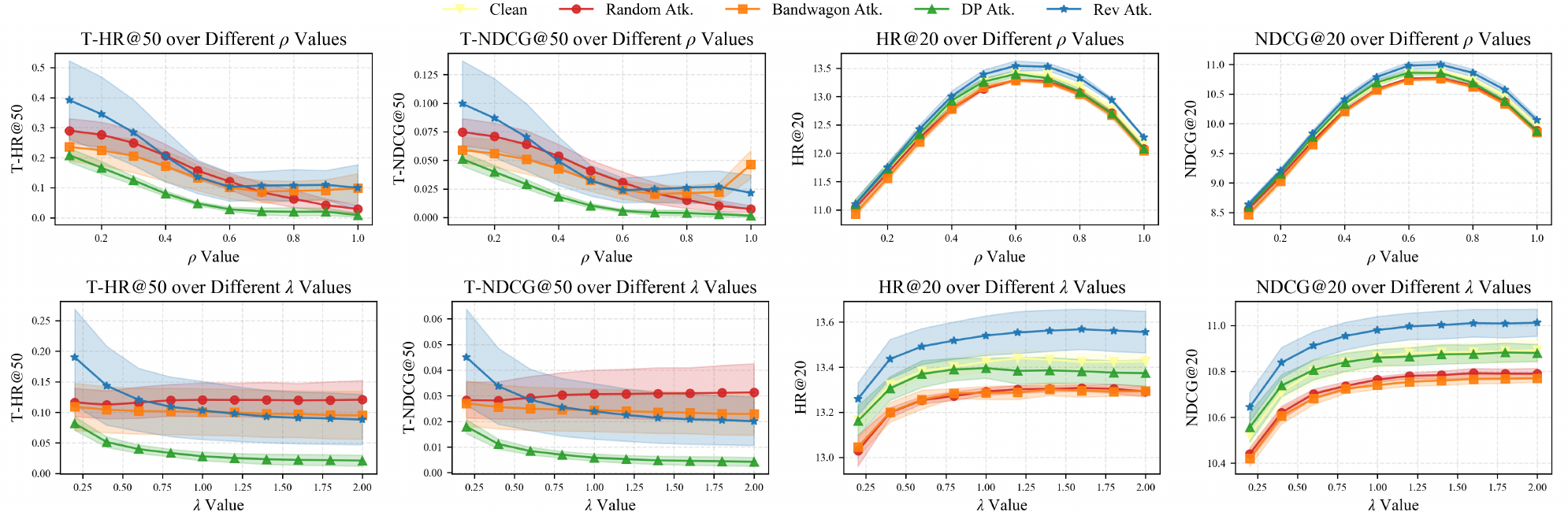}
    \caption{Top Section: Analysis of hyper-parameter $\rho$; Bottom Section: Analysis of hyper-parameter $\lambda$. (on Gowalla)}
    \label{fig:hyper}
\end{figure*}

In this section, we answer \textbf{RQ2}. We explore the effects of hyperparameters, i.e., perturbation magnitude $\rho$ and adversarial training weight $\lambda$ as defined in Equation~\ref{eq:vat}. The results are shown in Figure~\ref{fig:hyper}.

\subsubsection{Analysis of Hyper-Parameters $\rho$}
With $\lambda$ set to 1.0 (the optimal setting for $\lambda$), we vary $\rho$ from 0.1 to 1.0 in increments of 0.1. Our findings reveal a progressive improvement in defensive performance as $\rho$ increases, as shown in the top-right part of Figure~\ref{fig:hyper}. Notably, defensive efficacy stabilizes once $\rho$ exceeds 0.6. Furthermore, the range of $\rho$ from 0.5 to 0.8 results in the most significant enhancement in recommendation performance, as depicted in the top-right part of Figure~\ref{fig:hyper}. Even when $\rho$ is set at a small value of 0.2, there is a noticeable improvement in recommendation performance.

\subsubsection{Analysis of Hyper-Parameters $\lambda$}
Setting $\rho$ at 0.6 (the optimal setting for $\rho$), we vary $\lambda$ from 0.2 to 2.0 in increments of 0.2. This analysis shows that both defense capabilities and the ability to enhance recommendation quality become stable when $\lambda$ exceeds 1.0. Notably, excessively high $\lambda$ can increase performance variance and complicate the convergence of adversarial training. This suggests an optimal setting of $\lambda = 1.0$ to balance performance and stability.

\subsection{Case Study~(RQ3)}

\begin{figure}
    \centering
    \includegraphics[width=0.475\textwidth]{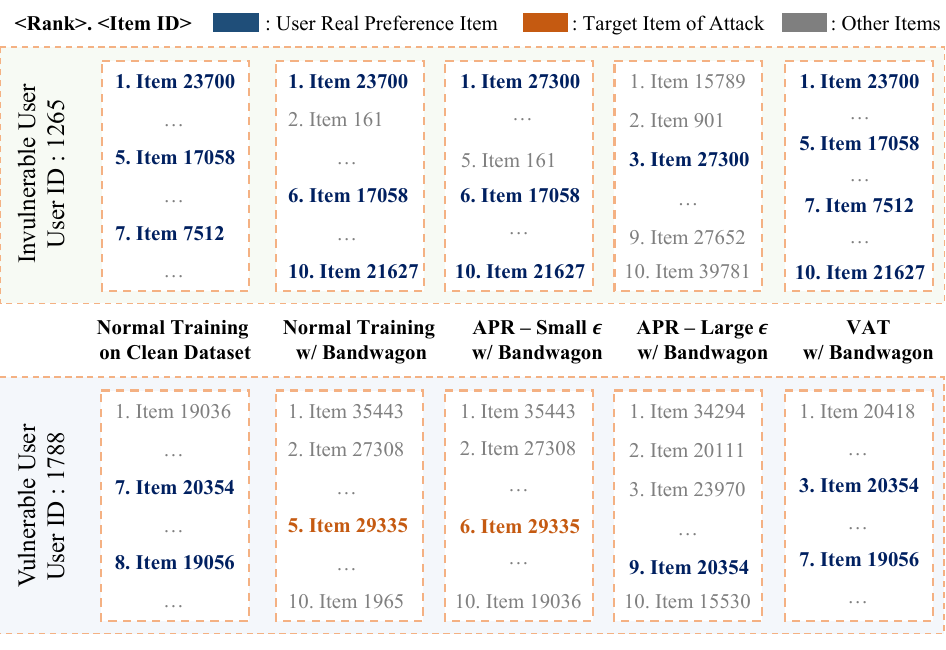}
    \caption{Cases of invulnerable user and vulnerable user.}
    \label{fig:user_case}
\end{figure}

\begin{figure}
    \centering
    \includegraphics[width=0.475\textwidth]{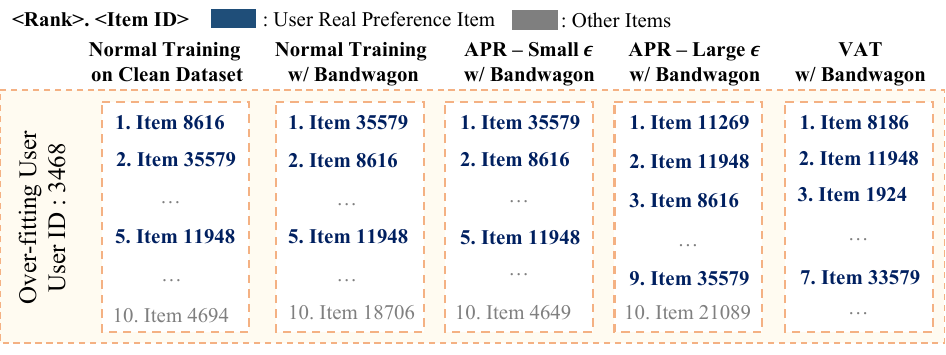}
    \caption{Case of over-fitted user.}
    \label{fig:user_case_of}
\end{figure}

In this section, we answer \textbf{RQ3} by presenting user cases that further support the effectiveness of VAT.

\subsubsection{Invulnerable User and Vulnerable User}
We illustrate cases for both an invulnerable user and a vulnerable user, showcasing their top-10 recommendation lists obtained through normal training on clean data, as well as through normal training, traditional adversarial training with the same small-magnitude perturbations ($\epsilon = 0.2$), and with the same large-magnitude perturbations ($\epsilon = 0.7$), and our VAT method on poisoned data as depicted in Figure~\ref{fig:user_case}.

We find that small-magnitude perturbations in traditional adversarial training preserve the recommendation performance for the invulnerable user (characterized by a large loss), but offer insufficient protection for the vulnerable user (characterized by a small loss). Conversely, large-magnitude perturbations in traditional adversarial training render the attack ineffective for the vulnerable user but impair the recommendation performance for the invulnerable user. With VAT, user-adaptive magnitudes of perturbations not only enhance recommendation performance for the invulnerable user but also provide adequate protection for the vulnerable user.

\subsubsection{Over-fitting User}
\label{sec:over_fit}
Additionally, we discuss users who are over-fitted (characterized by small losses but not affected) as mentioned in Section~\ref{sec:VAT-train}. Although these users are not affected by attacks, applying large-magnitude perturbations through VAT can mitigate over-fitting, thus improving their performance, as demonstrated in Figure~\ref{fig:user_case_of}.

\section{CONCLUSION}
In this study, we innovatively explore user vulnerability in recommender systems subjected to poisoning attacks. Our findings indicate that well-fitted users in the current recommender system are more likely to be vulnerable, i.e., affected by attacks. This exploration has led to the development of a Vulnerability-Aware Adversarial Training (VAT) method. VAT distinctively tailors the magnitude of adversarial perturbations according to users' vulnerabilities, thereby avoiding the typical trade-offs between robustness and performance suffered by traditional adversarial training methods in recommender systems. Through comprehensive experimentation, we have confirmed the effectiveness of VAT. VAT not only reduces the success ratio of poisoning attacks but also improves the overall recommendation performance.

\begin{acks}
This work is funded by the National Key R\&D Program of China (2022YFB3103700, 2022YFB3103701), the Strategic Priority Research Program of the Chinese Academy of Sciences under Grant No. XDB0680101, and the National Natural Science Foundation of China under Grant Nos. 62102402, 62272125, U21B2046. Huawei Shen is also supported by Beijing Academy of Artificial Intelligence (BAAI).
\end{acks}

\bibliographystyle{ACM-Reference-Format}
\bibliography{ref}


\end{document}